%% file: cipanp.sm.phenix.tex
\documentclass[12pt]{article}
\RequirePackage{lineno}
\usepackage{graphicx}
\usepackage{xspace}

\newcommand\pubdate{\today}

\def\vandy{Department of Physics and Astronomy\\
Vanderbilt University, Nashville, TN 37235, USA}
\def\support{\footnote{See acknowledgements of Refs. \cite{
Aidala:2017pup,Adare:2017wlc,Aidala:2018mcw} for support.}}

\def\Title#1{\begin{center} {\Large #1 } \end{center}}
\def\Author#1{\begin{center}{ \sc #1} \end{center}}
\def\Address#1{\begin{center}{ \it #1} \end{center}}

\newcommand\pubblock{\rightline{\begin{tabular}{l} 
         \pubdate  \end{tabular}}}
\newenvironment{Abstract}{\begin{quotation}  }{\end{quotation}}
\newenvironment{Presented}{\begin{quotation} \begin{center} 
             PRESENTED AT\end{center}\bigskip 
      \begin{center}\begin{large}}{\end{large}\end{center} \end{quotation}}

\input econfmacros.tex

\newcommand{\sqsn}{\mbox{$\sqrt{s_{_{NN}}}$}\xspace}

\newcommand{\dau}{\mbox{$d$$+$Au}\xspace}
\newcommand{\pau}{\mbox{$p$$+$Au}\xspace}
\newcommand{\hau}{\mbox{$^3$He$+$Au}\xspace}
\newcommand{\pdhau}{\mbox{$p/d/^{3}$He$+$Au}\xspace}

\newcommand{\nth}{\mbox{$n^{\rm th}$}\xspace}
\newcommand{\ampt}{{\sc ampt}\xspace}
\newcommand{\sonic}{{\sc sonic}\xspace}
\newcommand{\ssonic}{super{\sc sonic}\xspace}

\begin{document}
\begin{titlepage}
\pubblock

\vfill
\Title{PHENIX results on collectivity in small systems}
\vfill
\Author{Sylvia Morrow for the PHENIX Collaboration\support}
\Address{\vandy}
\vfill
\begin{Abstract}

PHENIX has measured elliptic and triangular 
flow of charged hadrons 
in \pdhau collisions at a center-of-mass 
energy of 200 GeV per nucleon pair,
and elliptic flow in \dau collisions at 
200, 62.4, and 39 GeV.
In order to asses the origin of collectivity in the smallest systems, 
these results are compared with several theoretical models that 
produce azimuthal particle correlations based on initial and/or 
final state effects. Hydrodynamical models, 
which include the formation of a 
droplet of quark-gluon plasma, provide the 
best simultaneous description of our observations. 
\end{Abstract}
\vfill
\begin{Presented}
Thirteenth Conference on the Intersections of Particle and Nuclear Physics\\
Palm Springs, CA, USA,  May 28 -- June 3, 2018
\end{Presented}
\vfill
\end{titlepage}
\def\thefootnote{\fnsymbol{footnote}}
\setcounter{footnote}{0}
%


%
%
%
%
\section{Introduction}\label{introduction}
Heavy-ion collision experiments, such as PHENIX, detect the 
particles produced when two nuclei collide 
at relativistic energies.
It has been established that there are many features
of this outgoing particle distribution which can 
be explained by the presence of a strongly
coupled fluid called a quark-gluon plasma (QGP).
There is a class of collision systems called ``small systems"
in which 
one or both of the colliding 
nuclei have very few nucleons.

An original motivation
for collecting small systems data was to provide a 
measurement of cold nuclear matter effects 
which could be contrasted with larger systems
which had a combination of hot and cold nuclear matter effects. 
However, among the fraction of small systems collisions
where the impact parameter was small, 
these data were found to have
produced particle distributions with features
that suggested the presence of QGP. 
As such, the origin of collectivity in small systems 
has been an ongoing debate.

These proceedings presents new PHENIX 
results in \pdhau collisions at \sqsn = 200 GeV
and additionally \dau collisions at lower energies 
(62.4 GeV and 39 GeV) 
\cite{Aidala:2017pup,Aidala:2016vgl,Adare:2015ctn}. 
The results show these systems can be described 
by hydrodynamical models.

%
%
%
%
\section{Experiment}\label{experiment}
The distribution of particles
collected by the PHENIX detector
during heavy-ion collisions
is influenced by many factors, including 
the presence (or absence) of QGP, 
the distribution of the partons or nucleons 
which participated in the collision, 
and the collision species. 
The azimuthal distribution of the produced particles
can be written as

\beq
   \frac{dN}{d\phi} \propto 1+ \sum_n 2 v_n\cos(n(\phi-\psi_n)),
\eeq{eq:dndphi}
where $v_n$ is the \nth order
coefficient to the expansion,
$\psi_n$ is the \nth order symmetry plane
angle determined on an event-by-event 
basis, and $\phi$ is the azimuthal 
angle of a detected particle.
We measure $\psi_n$ in the event plane
detector which is in a different region of phase space than
the particles of interest. 
This offers a simple observable in the form of the coefficients to
the Fourier expansion, $v_n$. 

In practice there are several methods for
calculating the coefficients. Here we show 
results using the event plane 
method \cite{Poskanzer:1998yz}
which finds $v_n$ to be
\beq
	v_n = \frac{\vev{ {\rm cos}(n(\phi
	- \psi_n)) }}{{\rm Res}
	(\psi_n)},
\eeq{eq:vnpt}
where Res($\psi_n$) is the event plane resolution
calculated using the three-subevent method.

%
%
%
%
\section{Results}\label{results}

\subsection{Beam energy scan}
\begin{figure}[htp]
  \begin{center}
    \includegraphics[width=1.0\linewidth,
    trim={3.2cm 8.5cm 3.2cm 8.5cm},clip]{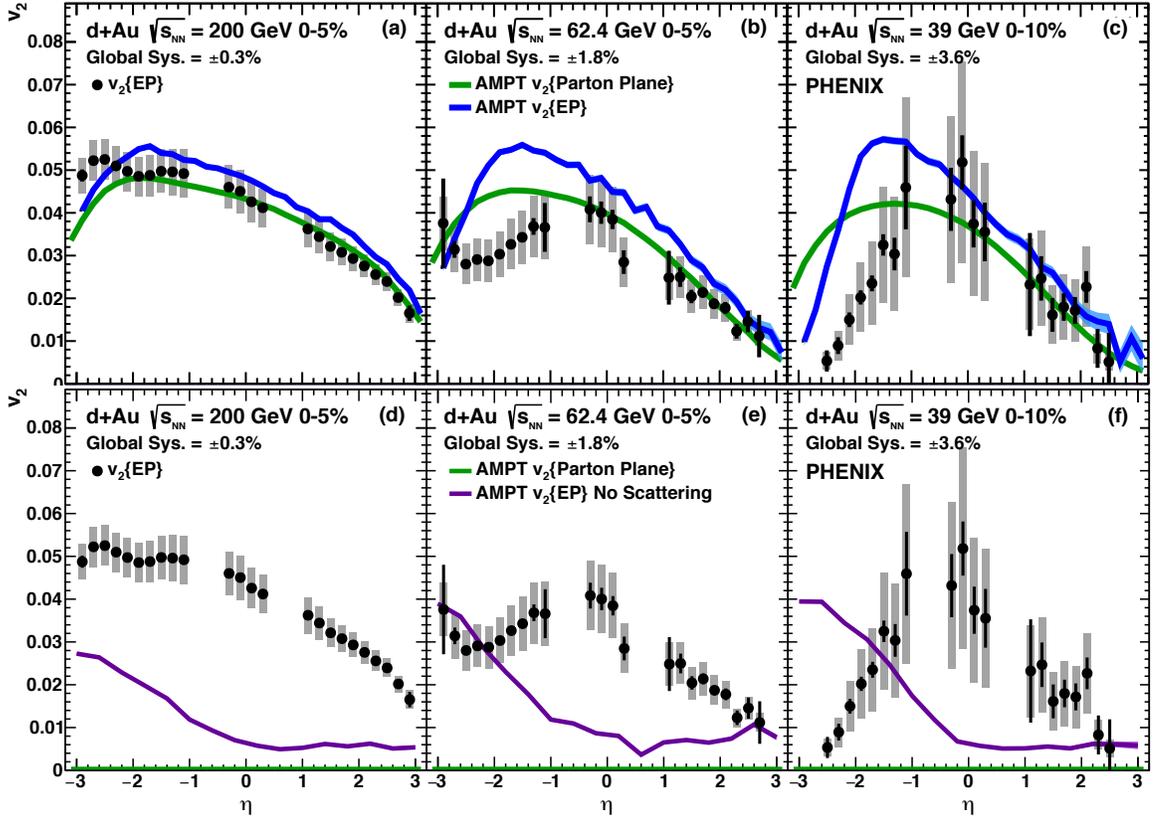}
    \caption{Measured $v_2(\eta)$ from \dau collisions at 
    		\sqsn = (a, d) 200 GeV, (b, e) 62.4 GeV, and (c, f) 39 
		GeV compared to \ampt calculations analyzed 
		using the parton plane (green line) and the 
		event plane (blue and purple lines).
		The top panels show the calculations
		with partonic and hadronic scattering,
		while the bottom panels have no scattering.
		Statistical (systematic) uncertainties are 
             	represented as vertical lines (boxes).}
    \label{fig:ppg205fig89}
  \end{center}
\end{figure}

In 2016, PHENIX collected data from \dau collisions
at \sqsn = 200, 62.4, 39, and 19.6 GeV. 
Fig. \ref{fig:ppg205fig89} shows measurements of the second order flow 
coefficient, $v_2$, from these top three energies measured as a function of 
the pseudorapidity, $\eta$ \cite{Aidala:2017pup}.
The shape of this distribution differs between 
energies in the region closest to the event plane
detector ($\eta<-3$). 
At backward rapidity ($\eta<0$)
the signal
is largest for \sqsn = 200 GeV. 
In contrast, the distribution at forward rapidity ($\eta>0$)
is roughly the same for all energies. 
At the lowest energy, $v_2(\eta)$ is roughly
symmetric so as the the collision energy 
increases and only one side of the distribution 
changes, the distribution becomes increasingly asymmetric.

The theory curves in Fig. \ref{fig:ppg205fig89}
are \ampt
calculations analyzed using the parton plane and the event plane.
The parton plane is 
calculated directly from the distributions of participating partons
in each event, so we describe it as a pure
flow signal. 
In contrast, the event plane calculation extracts this
distribution from the final state particle
distribution which is what we 
have experimentally accessible, 
offering a more direct comparison to the measurement. 
This calculation, therefore, includes 
any nonflow contributions which may enter the signal as a 
result of the analysis method limitations. 
As such, we consider this the
combined flow and nonflow signal. 

The top panels of Fig. \ref{fig:ppg205fig89} show 
that in forward rapidity regions, where 
the nonflow contribution to the measurement is small, the parton
plane calculation is in excellent agreement with the data, and the 
event plane calculation is reasonable though systematically a bit
too large. 
At backward rapidity the calculations diverge from the measurement,
though the event plane calculation may be qualitatively more similar.
More notably, in this rapidity regions there is a cross over between
the two \ampt calculations suggesting that the relationship between flow
and nonflow is not simply additive.

This feature is emphasized by the bottom panels of 
Fig. \ref{fig:ppg205fig89} which 
compares the measurement to 
\ampt calculations where both partonic
and hadronic scattering
has been turned off.
We see that the ``pure flow" as found using the 
parton plane is zero; therefore, 
the event plane calculation here is
exclusively nonflow. 
The value of $v_2\{{\rm EP}\}$ in panels d-f is largest 
at backward rapidity and cannot simply be added 
to $v_2\{{\rm Parton~Plane}\}$ in panels a-c
to yield the combined flow and nonflow calculation, 
$v_2\{{\rm EP}\}$ in panels a-c.

Further results from this beam energy scan include 
$v_2(p_T)$ at  \sqsn = 200, 62.4, 39, and 19.6 GeV 
\cite{Aidala:2017pup} and $v_2(N_{tracks})$ at the same four
energies \cite{Aidala:2017ajz}.

\subsection{Identified particle flow}
\begin{figure}[htbp]
\begin{center}
\includegraphics[width=0.3\linewidth]{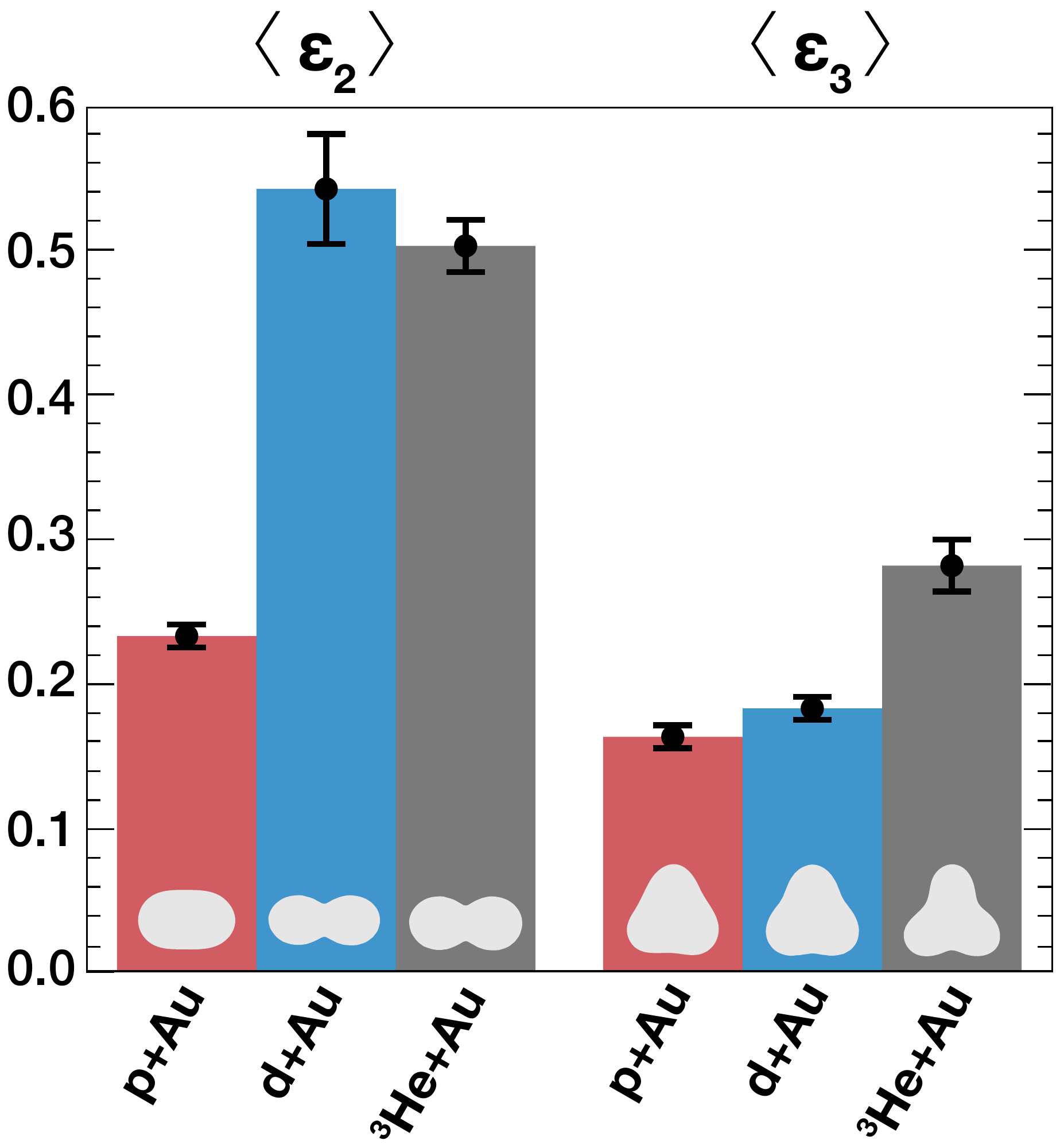}
\includegraphics[width=0.69\linewidth]{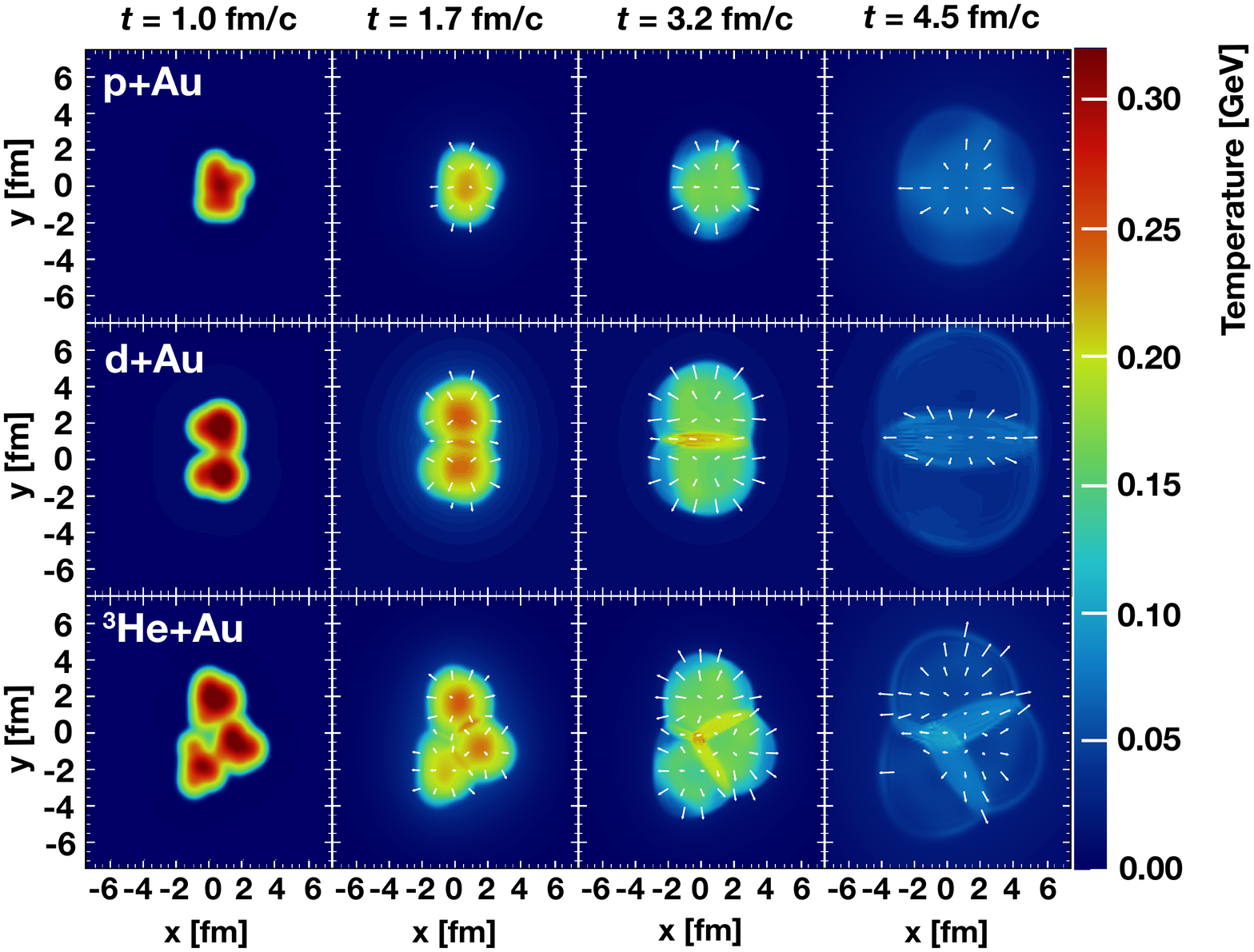}
    \caption{Average second and third order 
             eccentricity from Glauber Monte Carlo 
             simulations of the initial state for 
             central $p$/$d$/$^3$He+Au collisions (center) 
             with statistical uncertainties.
             Inset shapes illustrate the qualitative degree of 
             ellipticity ($\varepsilon_2$) or triangularity ($\varepsilon_3$).}
    \label{fig:ppg216fig1}
\end{center}
\end{figure}
%
\begin{figure}[htp]
  \begin{center}
    \includegraphics[width=1.0\linewidth]{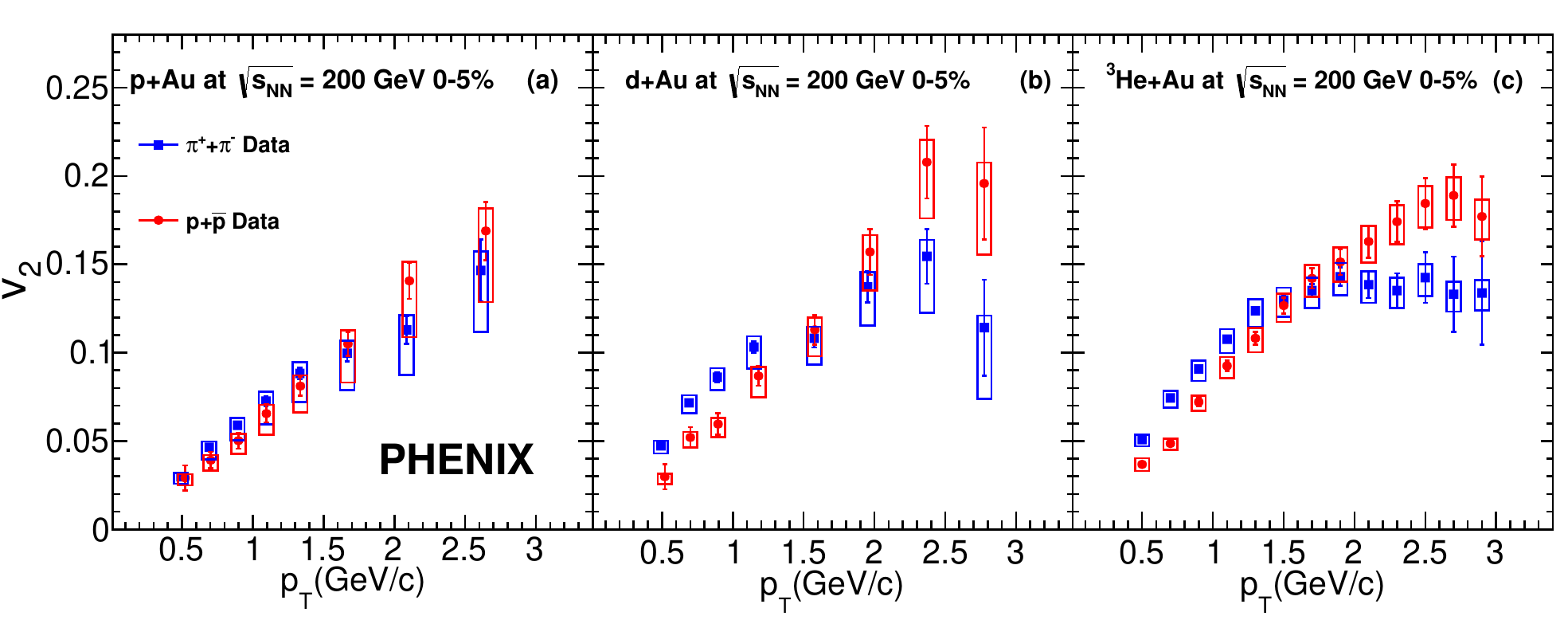}
    \caption{Measurements of $v_2(p_T)$ for identified pions 
             and protons in (a) $p$+Au, (b) $d$+Au, and 
             (c) $^{3}$He+Au collisions. 
             Statistical (systematic) uncertainties are 
             represented as vertical lines (boxes).}    
             \label{fig:ppg207fig2}
  \end{center}
\end{figure}
%
\begin{figure}[htp]
  \begin{center}
    \includegraphics[width=0.93\linewidth]{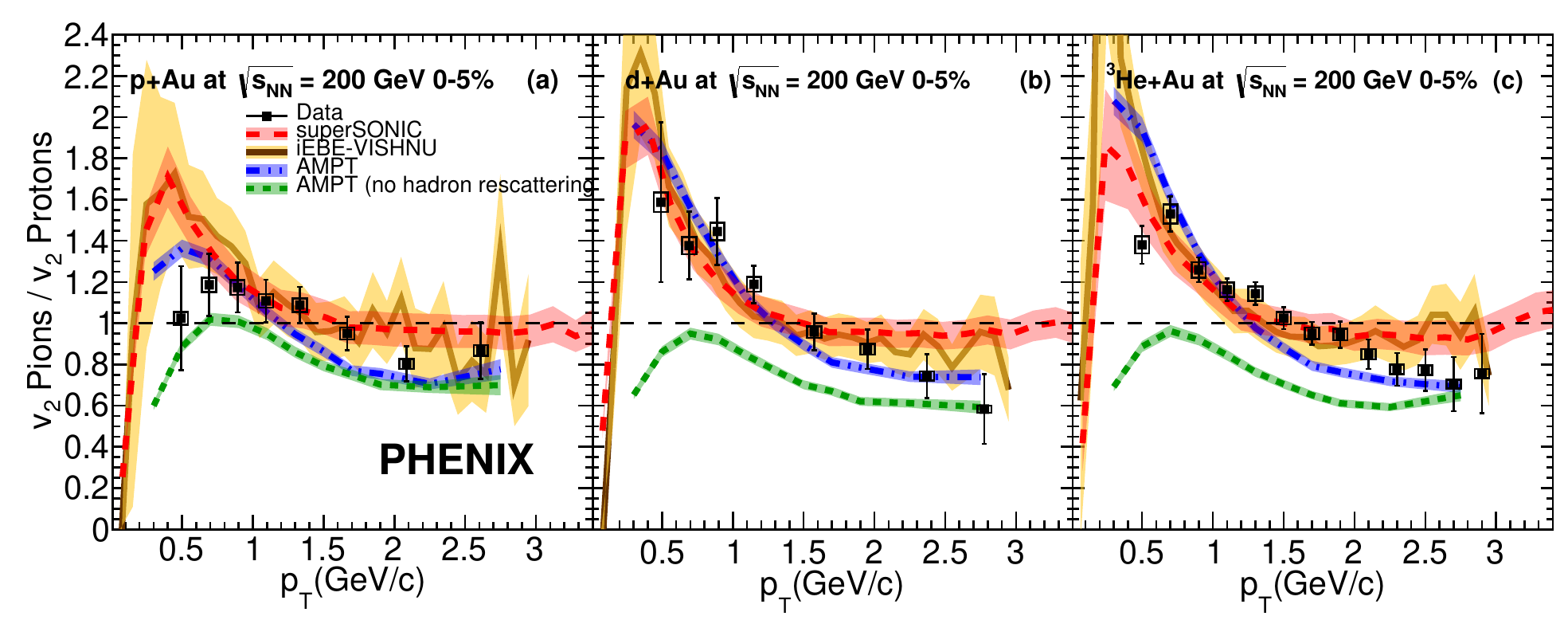}
    \caption{Measurements of pion $v_2$ over proton 
             $v_2$ as a function of $p_T$ in (a) $p$+Au, 
             (b) $d$+Au, and (c) $^{3}$He+Au collisions compared to 
             \ssonic, iEBE-VISHNU, and \ampt with 
             and without hadronic rescattering.
             Statistical (systematic) uncertainties are 
             represented as vertical lines (boxes).}
    \label{fig:ppg207fig4}
  \end{center}
\end{figure}

The second order flow 
coefficient, $v_2(p_T)$, was measured 
for identified pions and protons.
The shape of $v_2(p_T)$ is different depending 
on the mass of the particle
as well as the initial 
collision geometry seen by comparing 
\pau, \dau, and \hau collisions \cite{Adare:2017wlc}.
The left panel of 
Fig. \ref{fig:ppg216fig1} shows the 
shape of the average initial spatial distribution
of \pdhau collisions, quantified as \nth order eccentricity
terms where the cartoon insets demonstrate the degree of
ellipticity ($\epsilon_2$) and triangularity ($\epsilon_3$).

The right panel of Fig. \ref{fig:ppg216fig1}
displays the temperature evolution of a collision
system according to the hydrodynamical model
\sonic for a characteristic collision in each of the
three collision systems, \pdhau. 
As the hot medium cools the particle distribution
expands translating the initial spatial distribution
into a velocity or momentum distribution.

The measurement shown in Fig. \ref{fig:ppg207fig2} is 
consistent with the hydrodynamic picture of the hadrons
being part of a common velocity field.
Protons are much heavier than pions, so they 
have a different momentum distribution than 
pions in the same collision events. 
Across all three collision systems a general
pattern emerges: at low $p_T$ (below 
$p_T \approx 1.5$ GeV/c) the pion $v_2$
is larger than the proton $v_2$ and at high $p_T$ (above 
$p_T \approx 1.5$ GeV/c) 
the proton $v_2$ is larger than the pion $v_2$.

This trend can also be observed in
the ratio of the pion $v_2$ over
the proton $v_2$, as seen in Fig. \ref{fig:ppg207fig4}. 
The ratio emphasizes the differences 
between the distributions of the
two different hadrons and allows some uncertainty to cancel. 
Fig. \ref{fig:ppg207fig4} also compares these measurements
to several model calculations. 
The hydrodynamic models shown, 
\ssonic and iEBE-VISHNU, 
are in reasonable agreement
with the data at low $p_T$, 
but fail being completely flat above $p_T \approx 1.5$ GeV/c. 
In contrast, the \ampt calculations shown
are qualitatively better than the 
hydrodynamics calculations at
high $p_T$, but do not describe the low $p_T$ slope 
when hadronic rescattering is turned off. 
This feature suggests
that within the parton scattering 
framework, hadronic rescattering 
is essential to the formation of low $p_T$ mass splitting. 
This result also suggests that the difference in the treatment 
of hadronization is important to the formation of high $p_T$
mass splitting, as the hydrodynamical models use Cooper-Frye
prescription and \ampt uses quark recombination.

\subsection{Inclusive charged hadron flow}
\begin{figure}[htbp]
\begin{center}
\includegraphics[width=0.49\linewidth]{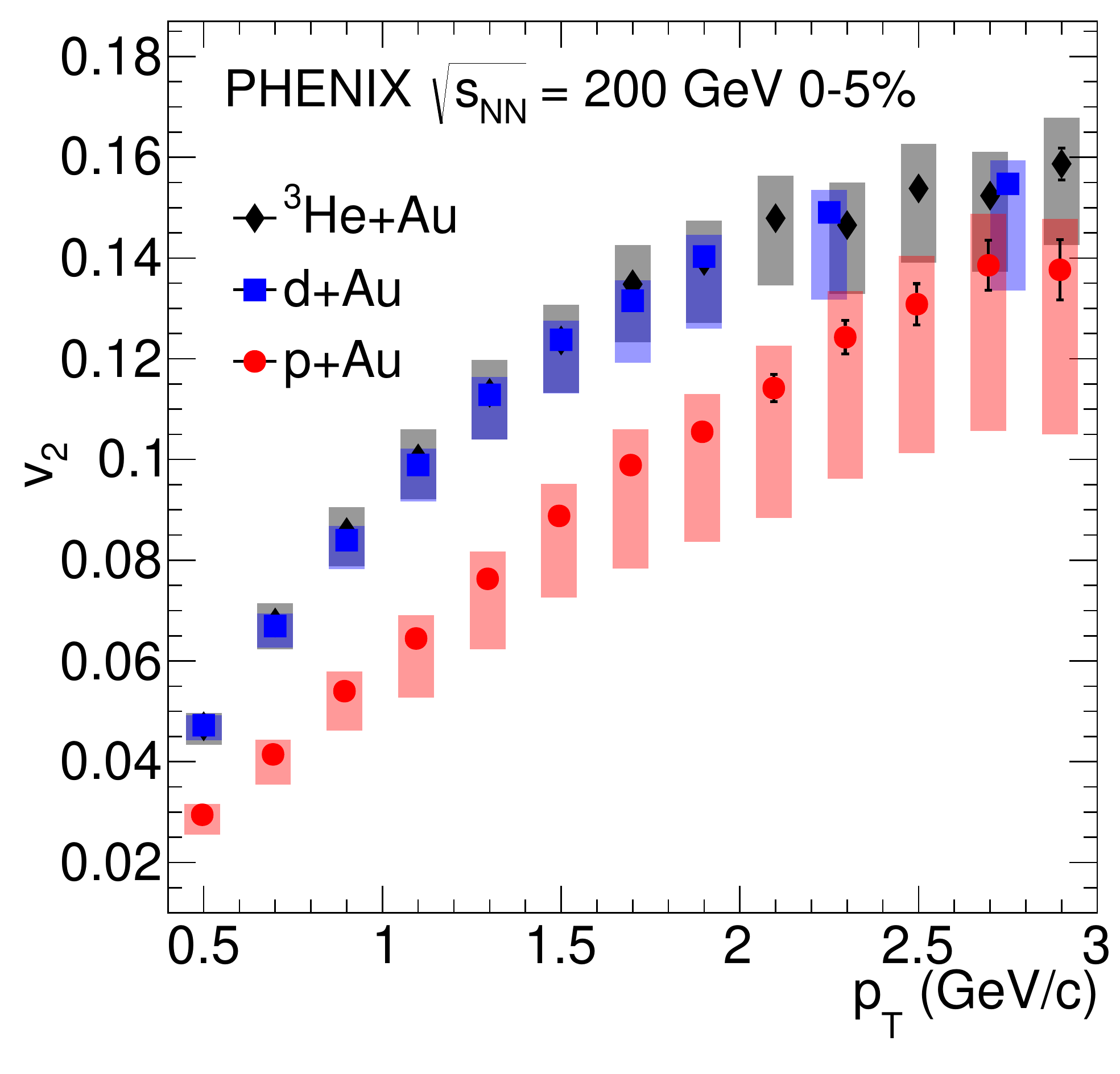}
\includegraphics[width=0.49\linewidth]{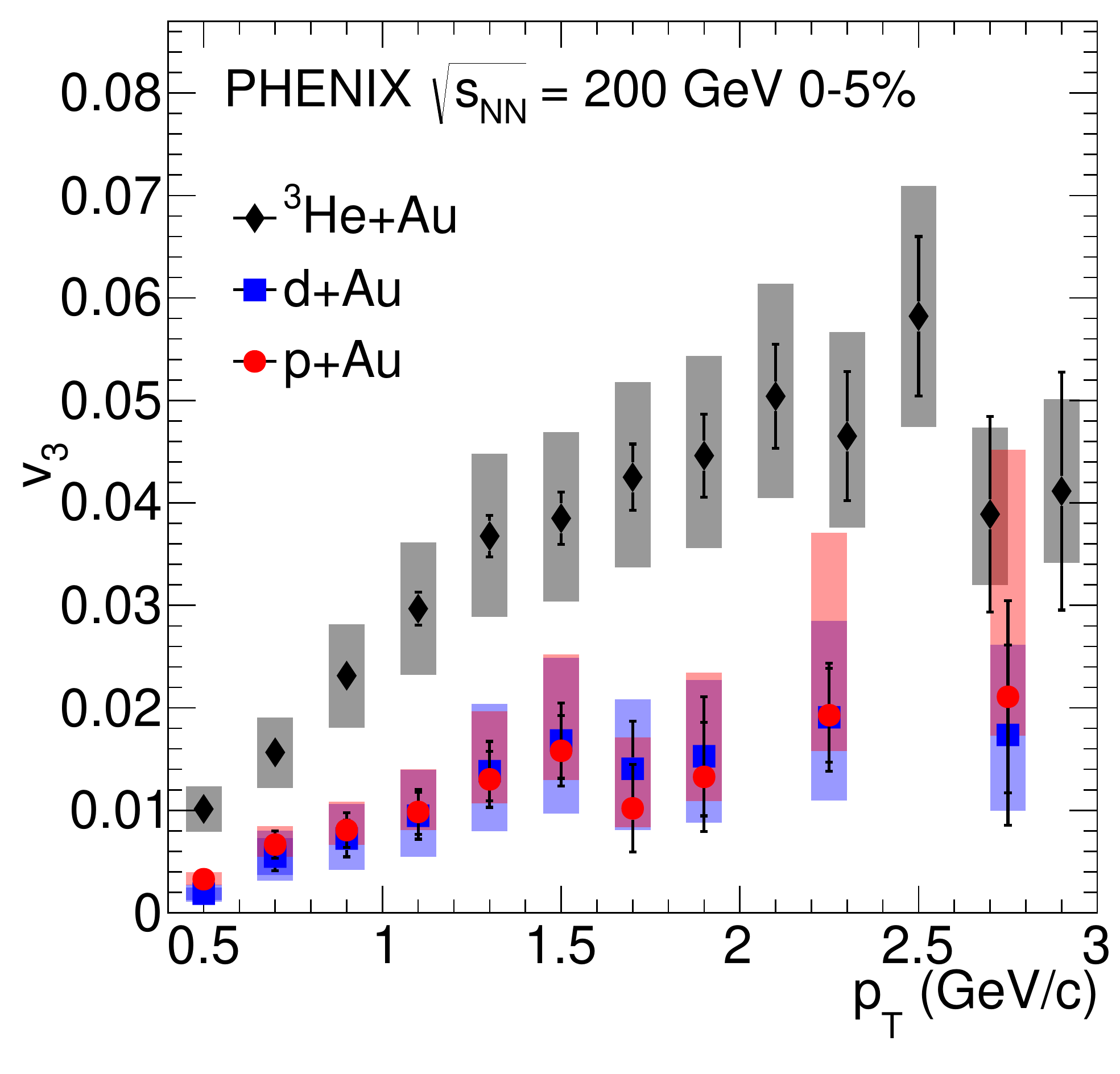}
    \caption{Measurements of $v_2(p_T)$ (left) and $v_3(p_T)$ (right) 
    	    in $^{3}$He+Au (black diamonds),
             $d$+Au (blue squares), $p$+Au (red circles).
             Statistical (systematic) uncertainties are 
             represented as vertical lines (boxes).}
    \label{fig:ppg216fig2}
\end{center}
\end{figure}

A measurement of the second and third
order flow coefficients for inclusive charged hadrons 
is shown in Fig. \ref{fig:ppg216fig2}. As was shown
in the left panel of Fig. \ref{fig:ppg216fig1}, there 
is a distinct ordering of $\epsilon_n$ across
the three collision systems, \pdhau. 
%
%
\begin{figure}[htp]
  \begin{center}
    \includegraphics[width=1.0\linewidth]{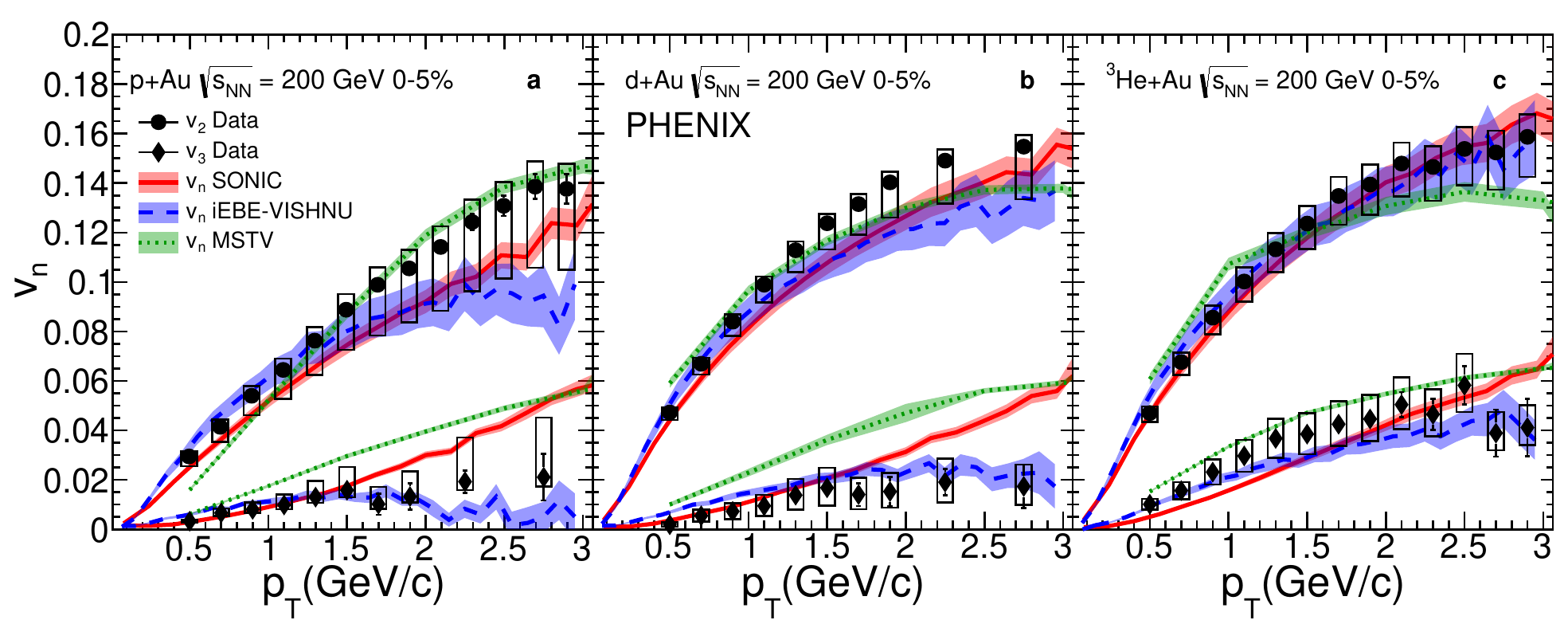}
    \caption{Measured $v_2(p_T)$ (circles) and $v_3(p_T)$ (diamonds)
             in (a) $p$+Au,  (b) $d$+Au, and (c) 
             $^{3}$He+Au collisions compared to hydrodynamical model 
             predictions from {\sc sonic} 
             (solid red), iEBE-VISHNU (dashed blue), and MSTV (dotted green).
             Statistical (systematic) uncertainties are 
             represented as vertical lines (boxes).}
    \label{fig:ppg216fig3}
  \end{center}
\end{figure}
For the second order, 
$\langle \epsilon_2 \rangle_{p+{\rm Au}} \approx \langle \epsilon_2 \rangle_{d+{\rm Au}} < \langle \epsilon_2 \rangle_{^3{\rm He+Au}}$ 
which matches the result shown in the left panel of Fig. \ref{fig:ppg216fig2},
$(v_2)_{p+{\rm Au}} \approx (v_2)_{d+{\rm Au}} < (v_2)_{^3{\rm He+Au}}$.
For the third order, 
$ \langle \epsilon_3 \rangle_{p+{\rm Au}} < \langle \epsilon_3 \rangle_{d+{\rm Au}} \approx \langle \epsilon_3 \rangle_{^3{\rm He+Au}}$
which matches the result shown in the right panel of Fig. \ref{fig:ppg216fig2},
$ (v_3)_{p+{\rm Au}} < (v_3)_{d+{\rm Au}} \approx (v_3)_{^3{\rm He+Au}}$.
This ordering is in agreement with a 
direct and efficient translation of the initial spatial 
distribution ($\epsilon_n$) to the final 
momentum distribution ($v_n$). 

For a more direct comparison, Fig. \ref{fig:ppg216fig3} 
shows the $v_2$ and $v_3$ measurements compared 
to calculations from several models. 
\sonic and iEBE-VISHNU,
which are both hydrodynamic models, are in good
agreement with the $v_2$ and $v_3$ measurements
in the three collision systems that have significantly
different initial geometries. 
MSTV is a postdiction
in the initial momentum correlation framework. While it 
has some success for $v_2$,
$v_3$ in \pau and \dau collisions are calculated to
be significantly too large.
Since the presentation of this talk 
at CIPANP, a detailed statistical analysis 
of the data to theory comparison has
been performed which
concludes that the 
hydrodynamics offers the best simultaneous 
description of the data \cite{Aidala:2018mcw}. 

%
%
%
%
\section{Conclusion}\label{conclusion}
These proceedings shows new PHENIX results
from small-systems collisions offering new 
insight into the relationship between the 
initial state and the final state and
how flow signals change with beam energy. 
It also highlights the complexity
of the topic and the interplay of 
different physics mechanisms.
The combination of all results shown here 
are best described by hydrodynamical models.

\newpage
\bibliographystyle{elsarticle-num}
\bibliography{cipanp.sm.phenix}

\end{document}

%% file: econfmacros.tex
\def\beq{\begin{equation}}
\def\eeq#1{\label{#1}\end{equation}}
\def\eeqn{\end{equation}}

\def\beqa{\begin{eqnarray}}
\def\eeqa#1{\label{#1}\end{eqnarray}}
\def\eeqan{\end{eqnarray}}

\let\bar=\overbar

\def\vev#1{\langle #1 \rangle}

\def\Dslash{\not{\hbox{\kern-4pt $D$}}}
\def\dslash{\not{\hbox{\kern-2pt $\del$}}}

\def\msb{{\bar{\ssstyle M \kern -1pt S}}}




%% file: cipanp.sm.phenix.bbl
\begin{thebibliography}{1}
\expandafter\ifx\csname url\endcsname\relax
  \def\url#1{\texttt{#1}}\fi
\expandafter\ifx\csname urlprefix\endcsname\relax\def\urlprefix{URL }\fi
\expandafter\ifx\csname href\endcsname\relax
  \def\href#1#2{#2} \def\path#1{#1}\fi

\bibitem{Aidala:2017pup}
C.~Aidala, et~al., {Measurements of azimuthal anisotropy and charged-particle
  multiplicity in $d$$+$Au collisions at $\sqrt{s_{_{NN}}}=$200, 62.4, 39, and
  19.6 GeV}, Phys. Rev. C96~(6) (2017) 064905.
\newblock \href {http://arxiv.org/abs/1708.06983} {\path{arXiv:1708.06983}},
  \href {http://dx.doi.org/10.1103/PhysRevC.96.064905}
  {\path{doi:10.1103/PhysRevC.96.064905}}.

\bibitem{Adare:2017wlc}
A.~Adare, et~al., {Measurements of mass-dependent azimuthal anisotropy in
  central $p$$+$Au, $d$$+$Au, and $^3$He$+$Au collisions at
  $\sqrt{s_{_{NN}}}=200$ GeV}, Phys. Rev. C97 (2018) 064904.
\newblock \href {http://arxiv.org/abs/1710.09736} {\path{arXiv:1710.09736}},
  \href {http://dx.doi.org/10.1103/PhysRevC.97.064904}
  {\path{doi:10.1103/PhysRevC.97.064904}}.

\bibitem{Aidala:2018mcw}
C.~Aidala, et~al., {Creating small circular, elliptical, and triangular
  droplets of quark-gluon plasma}~\href {http://arxiv.org/abs/1805.02973}
  {\path{arXiv:1805.02973}}.

\bibitem{Aidala:2016vgl}
C.~Aidala, et~al., {Measurement of long-range angular correlations and
  azimuthal anisotropies in high-multiplicity $p$$+$Au collisions at
  $\sqrt{s_{_{NN}}}=200$ GeV}, Phys. Rev. C95~(3) (2017) 034910.
\newblock \href {http://arxiv.org/abs/1609.02894} {\path{arXiv:1609.02894}},
  \href {http://dx.doi.org/10.1103/PhysRevC.95.034910}
  {\path{doi:10.1103/PhysRevC.95.034910}}.

\bibitem{Adare:2015ctn}
A.~Adare, et~al., {Measurements of elliptic and triangular flow in
  high-multiplicity $^{3}$He$+$Au collisions at $\sqrt{s_{_{NN}}}=200$ GeV},
  Phys. Rev. Lett. 115~(14) (2015) 142301.
\newblock \href {http://arxiv.org/abs/1507.06273} {\path{arXiv:1507.06273}},
  \href {http://dx.doi.org/10.1103/PhysRevLett.115.142301}
  {\path{doi:10.1103/PhysRevLett.115.142301}}.

\bibitem{Poskanzer:1998yz}
A.~M. Poskanzer, S.~A. Voloshin, {Methods for analyzing anisotropic flow in
  relativistic nuclear collisions}, Phys. Rev. C58 (1998) 1671--1678.
\newblock \href {http://arxiv.org/abs/nucl-ex/9805001}
  {\path{arXiv:nucl-ex/9805001}}, \href
  {http://dx.doi.org/10.1103/PhysRevC.58.1671}
  {\path{doi:10.1103/PhysRevC.58.1671}}.

\bibitem{Aidala:2017ajz}
C.~Aidala, et~al., {Measurements of Multiparticle Correlations in
  $d+\mathrm{Au}$ Collisions at 200, 62.4, 39, and 19.6 GeV and $p+\mathrm{Au}$
  Collisions at 200 GeV and Implications for Collective Behavior}, Phys. Rev.
  Lett. 120~(6) (2018) 062302.
\newblock \href {http://arxiv.org/abs/1707.06108} {\path{arXiv:1707.06108}},
  \href {http://dx.doi.org/10.1103/PhysRevLett.120.062302}
  {\path{doi:10.1103/PhysRevLett.120.062302}}.

\end{thebibliography}
